# MOOCs and crowdsourcing: Massive courses and massive resources
## by John Prpić, James Melton, Araz Taeihagh, and Terry Anderson

**Abstract**
Premised upon the observation that MOOC and crowdsourcing phenomena share several important characteristics, including IT mediation, large-scale human participation, and varying levels of openness to participants, this work systematizes a comparison of MOOC and crowdsourcing phenomena along these salient dimensions. In doing so, we learn that both domains share further common traits, including similarities in IT structures, knowledge generating capabilities, presence of intermediary service providers, and techniques designed to attract and maintain participant activity. Stemming directly from this analysis, we discuss new directions for future research in both fields and draw out actionable implications for practitioners and researchers in both domains.

**Contents**

**Introduction**

Recent research suggests that a growing proportion of formal education is now mediated by technology both inside and outside of traditional education institutions (Beaven, *et al.*, 2014). The rise of massive open online courses (MOOCs) exemplifies this trend. As the name implies, a MOOC is a form of IT-mediated education that represents a new mode of digital practice in formal education (Brown, *et al.*, 2014; Siemens, *et al.*, 2013; Weller and Anderson, 2013). As IT applications, MOOCc are differentiated from other IT-mediated formal education endeavours, such as learning management systems (Dalsgaard, 2006), by being completely open to the public at large, by being tuition-free for people to undertake formal learning, coalescing massive class enrollments, and explicitly drawing upon these massive class sizes to scale the education delivery itself. Beginning in 2008, MOOCs have seen a dramatic rise in prominence, with participant numbers for some individual courses reaching hundreds of thousands of people (Sinclair, *et al.*, 2015; Liyanagunawardena, *et al.*, 2013).

Almost simultaneous to the emergence of MOOCs, we have also witnessed the rise of crowdsourcing in the last decade (Brabham, 2008a). Crowdsourcing is an IT-mediated problem-solving, idea generation, and production modality, where problems and opportunities are broadcast through IT to the public at large, asking individuals to provide specific input for the problem or opportunity in question (Brabham, 2008b). Open calls serve to create IT-mediated crowds of individuals from the public at large, and in turn, these IT-mediated crowds can form in massive numbers, comprised of widely dispersed people (Prpić, *et al.*, 2015). Wikipedia is perhaps the most famous example of crowdsourcing (Prieur, *et al.*, 2008), though crowdsourcing also has many more focused implementations too, such as applications to policy-making (Prpić, *et al.*, 2015; 2014a; 2014b), health care (Prpić, 2015) public governance (Prpić and Shukla, 2014a), and private-sector innovation (Afuah and Tucci, 2012). In practitioner circles, the use of crowdsourcing as a productive tool for organizations has increased (Zhao and Zhu, 2014).

We observe that there are some fundamental similarities between MOOC and crowdsourcing phenomena. Both phenomena implement open calls to the public at large for participation, are solely IT-mediated phenomena, and form (Prpić, *et al.*, 2015; Padhariya and Raichura, 2014) and draw upon IT-mediated crowds for their existence and operation (Glassman, *et al.*, 2015; Li and Mitros, 2015; Mitros, 2015). Given these important similarities, it stands to reason that each field can learn something useful from a fine-grained analysis of the other.

Therefore, in this work we systematize these commonalities in order to undertake the first detailed comparison that we are aware of, spanning these otherwise distinct fields of research and practice. We first review the literature on MOOCs and organize the phenomena by delineating MOOCs from xMOOCs. From there, we review the literature on crowdsourcing, detailing the three generalized types of crowdsourcing. Next, we introduce literature from both domains that supports our comparison. We then combine the preceding frameworks into a table, comparing them along the fundamental dimensions that they share (IT, crowds, and openness). Before concluding, we discuss the ramifications of our analysis, illustrating the unique aspects of our contribution for both researchers and practitioners alike.

## MOOCs

MOOCs challenge the mainstream of formal education delivery and are experiencing exponential growth in the process of doing so (Saadatmand and Kumpulainen, 2014). Research indicates that the MOOC movement has arisen due to the proliferation of technology, increasing demand for educational opportunity, and shortcomings, notably cost and lack of access, of traditional formal education models (Saadatmand and Kumpulainen, 2014; Yuan and Powell, 2013). MOOCs have thus attracted the attention of educational institutions, teachers, course designers, politicians, policy-makers, researchers, entrepreneurs, and learners.

*The beginning of MOOCs*

The term 'MOOC' originated with Dave Cormier in 2008, in connection with a course at the University of Manitoba led by George Siemens and Stephen Downes that enrolled more than 2,000 students and employed multiple open educational resources in the form of IT tools such as wikis, online forums, Google Docs, YouTube, and Facebook groups to engage students and deliver the course (López-Sieben, *et al.*, 2014; Plasencia and Navas, 2014; Daradoumis, *et al.*, 2013). This pioneering MOOC operationalized a new pedagogical paradigm known as connectivism. Connectivism, presented as "a learning theory for the digital age" (Siemens, 2005), is in direct contrast to other learning paradigms such as cognitivism, constructivism, and behaviorism (Dron and Anderson, 2014; Saadatmand and Kumpulainen, 2014; Glance, *et al.*, 2013; Lane, 2009). Connectivism is an IT-mediated paradigm distinguished from the others by seeking to integrate emerging principles such as chaos, complexity, networks, and ubiquity into its explanation and prescriptions for formal education (Saadatmand and Kumpulainen, 2014; Dron and Anderson, 2014).

In addition to introducing the use of multiple, open IT tools to formal online education, Siemens and Downes also introduced peer review, peer assessment, and self-assessment notions into IT-mediated formal education delivery. These innovations in formal online education delivery enabled the original MOOC course to both create and accommodate an unprecedented scale of formal online education enrollment, and, as we shall see in the ensuing sections of this work, these innovations have been subsumed in whole or in part by the MOOCs that have come since.

In the time since these pioneering efforts, a variety of MOOC variants have evolved, and, perhaps unsurprisingly, a variety of taxonomies of MOOCs have also arisen in the research. For instance, some researchers now distinguish between cMOOCs (the original connectivist variety) from xMOOCs (extension MOOCs), typified by instructivist courses offered by Coursera, Udacity, and edX (Daniel, 2012). Whereas cMOOCs are tied to the new connectivist pedagogical approach, xMOOCs reflect "… a more traditional learning approach of knowledge duplication through video presentations and short quizzes and tests" (Saadatmand and Kumpulainen, 2014). On the other hand, Conole (2013) argues that the xMOOC/cMOOC distinction does not allow for quality design and instead she maps MOOCs to 12 different dimensions, including 'open,' 'massive,' 'degree of communication,' and 'degree of collaboration.' Similarly, Clark (2013) creates a taxonomy of MOOCs based on learning functionality. For example, he characterizes 'transferMOOCs,' which repurpose existing course content in a MOOC platform, 'synchMOOCs,' which have strict timelines, and 'asynchMOOCs,' which have an open timeline for course completion. From these studies we learn that MOOCs are still evolving rapidly. As the purpose of this paper does not require this level of distinction, we preserve and use the simpler xMOOC/cMOOC distinction.

*Evolving MOOCs*

Since their introduction in 2008, and the declaration by the *New York Times* that 2012 was the "Year of the MOOC", the MOOC terrain has continued to evolve. Some of the massiveness of early offerings has been lost in some cases, as enrollments are now more typically less than 10,000 students per course. As Jordan (2015) notes, "… enrollments on MOOCs have fallen while completion rates have increased" while completion rates, "… vary from 0.7% to 52.1%, with a median value of 12.6%". At the same time, the number of MOOC providers — of both content and IT-platforms — has increased. For example, the Class Central aggregator lists 334 MOOCs commencing in September 2015 alone.

Further, MOOCs have not and likely will not destroy traditional campus-based formal educational models, as predicted by some over-zealous early proponents. However, they have become an enduring and growing player in formal education, one that provides formal education alternatives for a massively large-scale of participants compared to traditional formal education, both face-to-face and online. One example of this growth is the use of MOOC content and IT-platforms by traditional educational institutions. In the next section, we detail this variety of MOOCs.

*cMOOCs and xMOOCs*

xMOOCs (Daniel, 2012) employ elements of the original MOOC, but are, in effect, branded IT platforms that offer content distribution partnerships to institutions. Glance, *et al.* (2013) note that the Courseras, Udacitys, and edXs of the world illustrate "… massive participation, online and open access, lectures formatted as short videos combined with formative quizzes, automated assessment and/or peer and self–assessment, and online forums and applications for peer support and discussion" (Daradoumis, *et al.*, 2013). Such xMOOCs have in many ways taken the innovative elements of the original connectivist MOOC detailed above, combining them into an integrated IT platform under one brand. One result of this integration has been that the scale of massive enrollment has increased by an order or magnitude or more, where the original connectivist MOOC counted more than 2,000 students total, HarvardX and MITx courses handle a cumulative and steady enrollment growth rate (over all the 68 courses offered at the time) of 2,200 participants per day (Ho, *et al.*, 2015).

However, xMOOCs also differ from classic connectivist MOOCs in crucial ways. Melton, *et al.* (2014) note that these kinds of courses, which they call third-party online courses, differ from classic MOOCs, because they are not always 'open', given that in many cases participation (or a certain level of participation) is restricted to students who have paid tuition while registered at a particular school. For this reason, these courses are often not massive, at least not in the sense of fostering a large community of students in one learning environment. Rather, in many cases xMOOCs function more like the traditional IT-mediated learning management systems (Blackboard or Moodle) that have been endemic in online higher education delivery for many years, though with the added ability to significantly scale delivery, allowing them to disseminate low-cost formal education content for third-party content providers such as schools and businesses (Savino, 2014; Anderson and McGreal, 2012). Given that some of the biggest names in academia (Stanford, MIT, Harvard, etc.) have founded xMOOCs (Ho, *et al.*, 2015), and in turn, provide branded formal educational content to other schools through the platforms, it may be that xMOOCs have the opportunity to cannibalize many traditional educational offerings, providing a disruptive factor in higher education (Stephens, *et al.*, 2015; Melton, *et al.*, 2014).

**Crowdsourcing**

The functions of crowdsourcing — problem-solving, idea generation, and production (Brabham, 2008a) — are achieved through different IT-mediated approaches. In the following subsections, we will describe the three types of crowdsourcing found in the literature (Prpić, *et al.*, 2015; de Vreede, *et al.*, 2013). Although these three categories of crowdsourcing are not necessarily exclusive or all-inclusive, they are a solid and reasonable basis upon which to undertake the aim of this paper, which is not to draw fine distinctions between the different types of crowdsourcing, as others before us have done (Prpić, *et al.*, 2015), but rather to examine the potential connections between the general kinds of crowdsourcing and cMOOCs/xMOOCs.

*Virtual labor markets*

A virtual labor market is an IT-mediated market for spot labor, exemplified by ventures like Crowdflower and Amazon's M-Turk. At these Web properties, workers agree to execute work in exchange for monetary compensation, and thus, these applications are thought to exemplify the 'production model' (Brabham, 2008b) of crowdsourcing. Workers undertake microtasks for pay, such as tagging photos, transcribing audio, and translating documents (Coetzee, *et al.*, 2014), and, in this way, human computation (Michelucci, 2013; Ipeirotis and Paritosh, 2011) is employed to undertake tasks that are not currently feasible for artificial intelligence to achieve. Microtasking through virtual labor markets can be rapidly completed, through massively parallel scale, with Crowdflower, for example, having over five million laborers available to undertake microtasks.

*Tournament crowdsourcing*

A separate form of crowdsourcing is known as tournament crowdsourcing. In tournament crowdsourcing, organizations post their problems or opportunities to IT-mediated crowds at fee-based Web properties, such as Innocentive, Eyeka, and Kaggle (Afuah and Tucci, 2012). In posting a problem or opportunity at the Web property, organizations create a prize competition, where all submitted entries are considered for awards, ranging from a total of a few hundred dollars, to a million dollars or more (Weller and Anderson, 2013). Tournament crowdsourcing Web properties generally attract and maintain specialized crowds of participants, premised upon the focus of the Web property. For example, the crowd at Eyeka coalesces around the creation of advertising collateral for brands, while the crowd at Kaggle forms around data science (Ben Taieb and Hyndman, 2014; Roth and Kimani, 2014). The crowds of participants at these sites is typically smaller when compared to virtual labor markets, where Kaggle, for example, has coalesced a crowd of about 140,000 data scientists to date (Prpić, *et al.*, 2015).

*Open collaboration*

In the open collaboration model of crowdsourcing, organizations post their problems and opportunities to the public at large through IT. Contributions from these crowds are voluntary and do not generally entail monetary exchange. Using social media applications (Crowley, *et al.*, 2014; Kietzmann, *et al.*, 2011) like Facebook and Twitter (Sutton, *et al.*, 2014) to garner contributions or starting an enterprise wiki (Jackson and Klobas, 2013), are primary examples of this type of crowdsourcing. The scale of open collaboration crowds can vary depending on the reach and engagement of the IT used, the efficacy of the open call for volunteers, and the degree of mass appeal of the call. Twitter, for example, has approximately 288 million registered users, and though this scale of crowd size is immense, there no guarantee that any significant subset of these potential contributors will pay attention to particular crowdsourcing efforts (Prpić, *et al.*, 2015).

*IT structure in crowdsourcing*

IT structure emanates from the crowd capital perspective (Prpić and Shukla, 2013; Prpić and Shukla, 2014b; Prpić, *et al.*, 2015; Prpić and Shukla, 2016), which generalizes the components and the dynamics of crowdsourcing (Kamerer, 2014; Massanari, 2012; Brabham, 2008a), prediction markets (Geifman, *et al.*, 2011), crowdfunding (Galuszka and Bystrov, 2014), open innovation platforms (Hallerstede, 2013; Frey, *et al.*, 2011), wikis (Mackey, 2011; Wilkinson and Huberman, 2007), and citizen science (Wiggins and Crowston, 2015) into a parsimonious model of IT-mediated crowds (Prpić and Shukla, 2013).

The crowd capital perspective informs us that heterogeneous knowledge resources can be generated through the organizational implementation of IT-mediated crowds. The generation of crowd capital is possible due to the existence of dispersed knowledge (Hayek, 1945) found in the individuals that comprise crowds. In addition, Prpić and Shukla (2013) make an important distinction in their model in respect to the types of IT used to engage crowds, where they distinguish between episodic and collaborative forms of IT for crowd engagement.

In episodic IT structures, the members of a particular crowd never interact with one another directly through the IT. A prime example of this type of IT structure is Google's reCAPTCHA (von Ahn, *et al.*, 2008), where Google accumulates significant knowledge resources (Palin, 2013), though it does so without any need for the crowd members to interact with one another. On the other hand, collaborative IT structures require that crowd members interact with one another through the IT for resources to form. Therefore, in collaborative IT structures, social capital must exist (or be created) through the IT for knowledge resources to be generated. A prime example of this type of IT structure is Wikipedia, where the crowd members build directly upon each other's contributions over time. This crucial distinction of IT structure materially impacts the form of the interface of the IT artifact used to mediate a crowd.

**The common elements of MOOCs and crowdsourcing**

As shown in the discussion above, the various types of crowdsourcing reveal relative differences in respect to the nature and size of the crowds that they attract, the level of openness that they display, and the IT structures that they implement. Similarly, in our review of the MOOC literature, we have learned that cMOOCs and xMOOCs illustrate stable, relative differences in respect to the scale of participants that they attract, the level of openness that they display, and the IT that they implement.

In the next section, we will use these stable and well-grounded similarities to undertake a comparison of the phenomena. However, before doing so, we further strengthen our use of these similarities by reviewing extant literature that explicitly combines MOOC and crowdsourcing phenomena.

*Crowdsourcing in formal education*

A small body of peer-reviewed literature exists, stemming predominantly from education researchers, that investigates the use of crowdsourcing in formal education (Al-Jumeily, *et al.*, 2015; Avery, 2014; Barbosa, *et al.*, 2014; Christensen and van Bever, 2014; Dontcheva, *et al.*, 2014; Dron and Anderson, 2014; Melville, 2014; Raman and Joachims, 2014; Clougherty and Popova, 2013; de Alfaro and Shavlovsky, 2013; Dow, *et al.*, 2013; Foulger, 2014; Kulkarni, *et al.*, 2013; Luger and Bowles, 2013; Recker, *et al.*, 2015; Solemon, *et al.*, 2013; Scalise, 2011; Skaržauskaitė, 2012; Weld, *et al.*, 2012; Anderson, 2011; Alario-Hoyos, *et al.*, 2013; Piech, *et al.*, 2013; Porcello and Hsi, 2013).

The majority of the literature describes crowdsourcing either as a method to create or aggregate educational resources for formal education (Al-Jumeily, *et al.*, 2015; Barbosa, *et al.*, 2014; Christensen and van Bever, 2014; Foulger, 2014; Dow, *et al.*, 2013; Recker, *et al.*, 2015; Solemon, *et al.*, 2013; Scalise, 2011; Skaržauskaitė, 2012; Weld, *et al.*, 2012; Anderson, 2011; Porcello and Hsi, 2013) or as a method toaid formal educational assessment (Avery, 2014; Kulkarni, *et al.*, 2013; Melville, 2014; Raman and Joachims, 2014; Clougherty and Popova, 2013; de Alfaro and Shavlovsky, 2013; Luger and Bowles, 2013; Weld, *et al.*, 2012; Piech, *et al.*, 2013).

This emerging corpus of research taken as a whole investigates the use of crowds in education in online, off-line, and blended formal educational settings. However, perhaps not surprisingly, given what we have illustrated thus far, the crowds implemented in these formal education settings are always IT-mediated, even when the formal education in question is not.

*Learning at scale*

At the same time, a new and emerging body of literature, stemming predominantly from computer science and HCI researchers, is investigating Learning at Scale (Glassman, *et al.*, 2015; Lasecki, *et al.*, 2015; Li and Mitros, 2015; Mitros, 2015; Mustafaraj and Bu, 2015; Williams, *et al.*, 2015a; Williams, *et al.*, 2015b; Veeramachaneni, *et al.*, 2015; Zhou, *et al.*, 2015; Asuncion, *et al.*, 2014; Chung, *et al.*, 2014; Dontcheva, *et al.*, 2014; Gillani, *et al.*, 2014; Kim, *et al.*, 2014; Mitros and Sun, 2014; Padhariya and Raichura, 2014; Williams, *et al.*, 2014; Nickerson, 2013).

This body of work is concerned solely with IT-mediated learning in both formal and informal education settings. Much of the work is focused solely upon formal education settings at xMOOCs in particular (Glassman, *et al.*, 2015; Li and Mitros, 2015; Mitros, 2015; Mustafaraj and Bu, 2015; Williams, *et al.*, 2015a; Williams, *et al.*, 2015b; Veeramachaneni, *et al.*, 2015; Zhou, *et al.*, 2015; Gillani, *et al.*, 2014; Mitros and Sun, 2014; Williams, *et al.*, 2014), while others are focused on informal education settings such as in corporations (Asuncion, *et al.*, 2014), or in crowdsourcing endeavors (Lasecki, *et al.*, 2015; Williams, *et al.*, 2015a; Chung, *et al.*, 2014; Dontcheva, *et al.*, 2014; Kim, *et al.*, 2014; Padhariya and Raichura, 2014; Nickerson, 2013).

As this latter example illustrates, this research explicitly connects learning and crowds (Gillani, *et al.*, 2014), and, further, the research acknowledges that resources of various kinds can be generated by IT-mediated crowds assembled for education purposes, in what is termed as 'learnersourcing' in this literature (Glassman, *et al.*, 2015; Li and Mitros, 2015; Mitros, 2015).

Yet, in all the works in this section, the corpus of literature does not distinguish between the similarities or differences among the different forms of crowdsourcing as they pertain to formal education, cMOOCs, and xMOOCs. In these works, crowdsourcing is generally treated as a singular phenomenon, usually focused either on open collaboration or virtual labor markets in the respective papers. Yet, as we have seen in this work thus far, crowdsourcing is not a singular phenomenon, and the differences between the types are both stable and important. Thus, much work is needed in the education, cMOOC, and xMOOC domains to acknowledge the extant crowdsourcing literature and to approach crowdsourcing in education with much more nuance and perhaps much more potential too.

*Social media in formal education*

Similar to the above literatures, numerous researchers have investigated the use and implementation of social media in formal education. Although it is very much beyond the scope of this work to review the entirety of this burgeoning literature, as others have done (Dron and Anderson, 2014; Selwyn, 2012; de Waard, *et al.*, 2011), for our purposes it is useful to point out two salient features of this literature.

First, in respect to formal education research investigating social media in MOOCs, most if not all of the research is focused on social media use in cMOOCS (de Waard, *et al.*, 2011), which, given our earlier analysis of MOOCs, is perhaps unsurprising. Second, as we illustrated earlier, social media use is considered to be a fundamental element of crowdsourcing. Likewise, our characterization of cMOOC and xMOOC participants as crowds is very well supported, as evidenced by the literature on open collaboration, the study of social media in formal education, and the learning at scale literature.

*Techniques to attract and maintain crowd activity*

Literature has emerged that investigates the use of reputation systems in formal education (Attali and Arieli-Attali, 2015; Buckley and Doyle, 2014; Caponetto, *et al.*, 2014; Coetzee, *et al.*, 2014), in xMOOCs (Krause, *et al.*, 2015; Vaibhav and Gupta, 2014), and in classic MOOCs (Gené, *et al.*, 2014). Similarly, there are also studies investigating the use of public award systems known as 'digital badges' in formal education (Abramovich, *et al.*, 2013; Goligoski, 2012). At the same time, there is a parallel body of literature investigating gamification techniques — which can include leaderboards, reputation systems, points, 'voting up', 'likes', etc. — in crowdsourcing (Kacorri, *et al.*, 2014; Tan, *et al.*, 2013; Eickhoff, *et al.*, 2012).

Taken together, these literatures would seem to indicate that similar techniques are already being used with both of these forms of IT-mediated crowds to engage individuals, and to maintain their participation.

*Summary*

Altogether, the entirety of the literature reviewed thus far indicates that there are significant, fundamental, and salient overlaps in MOOC and crowdsourcing phenomena, therein strongly supporting the fundamental observation of this work. However, these significant commonalities (IT mediation, crowds, and openness) have not been operationalized into a systematic framework to allow a more fine-grained picture of these important commonalities. Addressing this gap is the focus on the next section.

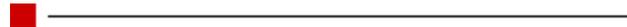

**Systematic analysis of MOOC and crowdsourcing phenomena**

In this section, we integrate the review of the MOOC and crowdsourcing literatures, and analyze the different forms of MOOCs and crowdsourcing in Table 1.

| | **Table 1: Comparison of types of MOOCs and types of crowdsourcing along common dimensions.** Note: https://www.duolingo.com/comment/862650 — Though most cMOOCS number in the thousands of participants or less. | | | |
|---|---|---|---|---|
| | **IT structure** | **Openness** | **Largest crowd size** | **Nature of crowd** |
| **Virtual labor markets** | Episodic | Private | Millions | General |
| **Open collaboration** | Collaborative | Public | Hundreds of millions | General |
| **Tournament crowdsourcing** | Episodic | Private | Hundreds of thousands | Specialized |
| **cMOOCs** | Episodic and collaborative | Public | Millions* | Specialized |
| **xMOOCs** | Collaborative | Public and private | Hundreds of thousands | Specialized |

*IT structure implemented*

From the comparison in Table 1, we can see that cMOOCs (in most cases), xMOOCs, and open collaboration crowdsourcing share similar collaborative IT structures. In these endeavours, crowd participants interact with one another through an IT interface, illustrating that social capital is an important common requirement amongst these endeavours. For instance, the use of Twitter exemplifies a collaborative IT structure in the open collaboration crowdsourcing domain, where the inherent social network of the application impacts both the quantity and quality of interaction by crowd members and thus the resources possible from such crowds. Similarly, in classic connectivist MOOCs, crowd-members must "connect" with one another in some form or another for the formal education to manifest. In regard to xMOOCs, it is well-known that peer assessment (Kulkarni, *et al.*, 2013; Raman and Joachims, 2014), group activities (Collazos, *et al.*, 2014), and reputation systems (Coetzee, *et al.*, 2014) entail direct individual collaboration within these crowds.

On the other hand, from Table 1, we can also see that cMOOCs, in some cases, along with virtual labor markets and tournament crowdsourcing, share similar episodic IT structures that do not necessitate the direct interaction of crowd participants through the IT. For example, in a virtual labor market, like Amazon's M-Turk, microtasks are undertaken independently by individual crowd participants. Similarly, in tournament crowdsourcing applications, like eYeka, the contest submissions of individual crowd participants are not available for review by other crowd members. In cMOOCs, like Duolingo (Garcia, 2013; von Ahn, 2013; Savage, 2012), crowd participants do not directly interact with one another through the IT. Thus, Duolingo is, in our view, a prime example of a MOOC with an episodic IT structure.

*Openness*

From the comparison in Table 1, we can see that cMOOCs, open collaboration crowdsourcing, and some xMOOCs are considered public, while tournament crowdsourcing, virtual labor markets, and some xMOOCs are considered private, with respect to openness. This distinction highlights the accessibility of the IT application to the public at large. In this respect, 'public' indicates that the application is free of charge for an individual or organization to use, while 'private' indicates that some cost is involved to use the application.

Individuals or organizations must pay to launch a competition at a tournament crowdsourcing site, like Innocentive, or to access the spot labor at virtual labor markets like Crowdflower. Similarly, as Melton, *et al.* (2014) point out, xMOOCs may charge for their services, thus making those courses private in nature. On the other hand, cMOOCs, such as Duolingo, are gratis to participate. In a similar fashion, open collaboration crowdsourcing is voluntary, and requires no monetary exchange to participate.

*Nature of the crowd*

From the comparison in Table 1, we can see that open collaboration crowdsourcing and virtual labor markets rely on what may be called general crowds, while tournament crowdsourcing, cMOOCs, and xMOOCs rely on specialized crowds. In this respect, specialized crowds can form around specific types of content, while general crowds do not.

For example, an individual using Duolingo to learn Spanish is a member of a specialized crowd interested only in Spanish and not necessarily all languages. In juxtaposition, participants at open collaboration crowdsourcing endeavors, such as Wikipedia or Twitter, form around multiple content types. The specialized or general nature of a crowd has important ramifications for the size of the potential crowd that is available to the endeavor, while also impacting the features of the IT used, tasks assigned to participants, and chosen pedagogy.

*Size of the crowd*

From the comparison in Table 1, we can see the largest known crowd size for each IT application. We see that the size of IT-mediated crowds represented by the applications reviewed here range from thousands of participants, to hundreds of millions.

Crowd size is an important factor for each of these applications, given that each application relies on their assembled crowd to deliver promised functionality. At the same time, crowd size serves as an upper limit to the scale of resources that can be created in each setting and also potentially the speed at which these resources can be generated (Prpić, *et al.*, 2015).

**Discussion**

The comparison of the common characteristics earlier reveals that the particular variants of cMOOCs and xMOOCs examined do not exactly mirror the three forms of crowdsourcing as noted earlier. In this respect, each form of IT displays a set of unique traits, while at the same time sharing common features across the entire pool of IT applications investigated here. This uniqueness within applications, yet commonalities across applications, is important to understand for both the research and practitioner communities.

For researchers, the across-application commonalities indicate that there is likely to be extant literature from the other fields that will be relevant to any of the five particular sub-domains investigated here. For instance, as we have illustrated, cMOOCs and xMOOCs both leverage crowds in their operation; therefore, the education and education technology literatures should stand to gain from the computer science, HCI, and MIS literatures in respect to the conception and operationalization of crowds. Doing so raises new and interesting questions in education research, where, for example, researchers could ask 'What is the effect of IT structure on learning efficacy in MOOCs?'

For practitioners, there is value in investigating these commonalities too: to gain ideas for the design, development, and administration of crowd-engaging IT. For instance, the algorithms used in xMOOCs for grading and assessment (Wu, *et al.*, 2015; Kwon and McMains, 2015;

Krause, 2014) may be able to shed light on the validation of microtask completion by individuals at virtual labor markets, where individuals are rated upon their historical performance with tasks.

*Virtual labor markets for formal education assessment*

When we look at the potentialities of the three types of crowdsourcing, we see interesting avenues for application to cMOOCs and xMOOCs, as well as to traditional education settings. For example, can human computation potentials found in virtual labor markets be applicable to cMOOCs, xMOOCs, and education delivery in general? Given our discussion of the learning at scale literature, and given that assessment is an endemic feature of education delivery in all forms, is it possible that virtual labor market crowds can be used to undertake formal education assessments quickly, cheaply, and effectively?

For example, already existing rubrics could be transposed into microtasks, or one entire microtask, to be put to virtual labor market crowds for evaluation. Then, given that virtual labor markets allow the massively parallel undertaking of tasks at low cost, virtual labor market evaluation of student work could provide almost instant assessment feedback. Though some may doubt a crowd's ability to render accurate assessments, the research indicates that in some very complicated venues a crowd can perform as well or better than experts (Lee, 2013; Mitry, *et al.*, 2013; Mortensen, *et al.*, 2013).

Similarly, given that numerous assessments can be received for each task (*i.e.*, task-duplication) and that each microtask worker is independent of one another, a tenable assessment could be achieved through the simple aggregation of the task results (Goligoski, 2012). Further, an added benefit of virtual labor market crowds is the possibility to choose a particular subset of the crowd to suit specific needs; for example, segmenting participants by geographic areas or by historical performance ratings. Further, virtual labor markets can be an order of magnitude more 'massive' than cMOOCs and xMOOCs (Prpić, *et al.*, 2015), thus providing a wealth of potential capability to service cMOOC and xMOOC assessment needs.

It is worth emphasizing that we are not advocating for nor predicting the elimination of the teaching assistant (TA) or professor in any way. On the contrary, the proper use of crowds for certain types of formal education evaluation could free the TA or the professor to engage with students in higher-value-added activities, such as mentoring, holistic evaluation, and discussion. This may allow for the maintenance and perhaps even the enhancement of learner-teacher educational relationships. These changes could function similarly to the advent of the textbook, which outsourced some of the professor's traditional role of content transmission. In any case, as Harris and Srinavasan (2012) illustrate, although professors may not be using crowds for educational purposes, students may already be doing so for assignments.

*Knowledge and learning*

Another outcome of our investigation of MOOCs and crowdsourcing is the knowledge-based by-products of IT-mediated crowds. As mentioned earlier, both connectivist MOOCs and xMOOCs (through 'learnersourcing') are thought to generate knowledge with dispersed learners. Likewise, as discussed earlier, crowdsourcing is seen by many (Brabham, 2008a; Prpić and Shukla, 2014a) as a knowledge-generation activity. Thus, it would seem that the more 'overt' knowledge generation activities in cMOOCs and xMOOCs could widely inform crowdsourcing research and practice.

For instance, consider that in virtual labor markets, there is often little or no sharing, archiving, or leveraging of the post-task knowledge gained from individual task-work experience. Though isolation is perhaps necessary in some task performance, it is not conducive to learning and knowledge sharing. In fact, Amazon's M-Turk workers have self-organized outside of the M-Turk platform to share information (Lease, *et al.*, 2013). Further, although machine learning is being studied in the setting of virtual labor markets (Quinn, *et al.*, 2009; Quinn and Bederson, 2011), little research (Lasecki, *et al.*, 2015; Williams, *et al.*, 2015a; Chung, *et al.*, 2014; Dontcheva, *et al.*, 2014; Kim, *et al.*, 2014; Padhariya and Raichura, 2014; Nickerson, 2013) has investigated the learning (machine or individual) of participants in virtual labor markets, open collaboration or tournament crowdsourcing writ large.

Is it possible that virtual labor markets can apply the knowledge generation activities of xMOOCs or episodic MOOCs like Duolingo to facilitate crowdsourcing participant learning? Given that the research (Hacker, 2014; Michelucci, 2013; Ipeirotis and Paritosh, 2011) informs us that both Duolingo and virtual labor markets, for example, manifest human computation from large IT-mediated crowds, it should be possible for learning to be imbued in virtual labor market processes. In theory, there should be little difference in the learning capabilities of IT-mediated workers at virtual labor markets and IT-mediated learners in cMOOCs and xMOOCs.

*Intermediation and analytics*

Additionally, this work illustrates that crowdsourcing and MOOCs have each spawned numerous intermediary service providers. While xMOOCs, as a separate class of MOOCs, are defined by intermediation, each of the three forms of crowdsourcing investigated here also displays many examples of intermediaries providing crowd-engaging services.

In the virtual labor market field, in particular, a second level of intermediary service provider is now emerging — companies like Crowdsource or EnableVue — who supply services to help organizations prepare tasks for virtual labor market crowds and then engage the virtual labor market crowds for the client as well. One wonders how long it will be until a second level of xMOOCs emerges, perhaps offering "pick and choose" content from numerous cMOOCs, and other xMOOCs in one IT setting.

On a related note, learning analytics and educational data mining have now emerged as a major new field of inquiry (Sin and Muthu, 2015), where the emergence of big data from new educational IT, combined with advances in computation, holds promise for improving learning processes in formal education and beyond (Siemens and Baker, 2012). These new potentialities, driven by digital trace data (Yoo, *et al.*, 2012) and also from the emergence of cMOOCs and xMOOCs, represent a 'fundamental shift in how education systems function' (Siemens and Baker, 2012). Given that this data and related analytics are aimed at assessment and appropriate learning interventions that inform both humans and algorithms, we see the learning analytics movement as being able to assist crowdsourcing providers in rating and supporting individual crowd participants as well.

## Conclusion

In this paper, we reviewed the literature on MOOCs and crowdsourcing and considered their various forms and characteristics. Then, we compared MOOCs and crowdsourcing and unpacked their similarities and differences vis-à-vis the IT-structure utilized, their relative levels of openness, and the types and size of crowds that they engage.

In respect to the crowdsourcing literature, we advance this literature by providing a comparison of crowdsourcing types across three universal dimensions: IT structure, openness, and crowd type. This approach serves to highlight important similarities, differences, and trade-offs of crowdsourcing modalities. At the same time, we single out areas of research and practice within the cMOOCS and xMOOCs literature that may provide useful knowledge for crowdsourcing researchers and practitioners to investigate further.

For education and online learning literature, we highlight important similarities and differences for cMOOCs and xMOOCs along three dimensions: IT structure, openness, and crowd type. We also aggregate and describe a corpus of emerging literature that investigates crowdsourcing in education, learning at scale, and learnersourcing, and discuss how some features of crowdsourcing applications could be implemented in cMOOC/xMOOC domains. In achieving these ends, we also suggest how crowdsourcing applications might be implemented in conjunction with cMOOCs and xMOOCs.

Further, we note that the MOOC and crowdsourcing fields are both emerging rapidly, and that many applications in each field are likely to defy easy categorization. Therefore, our work here should not be considered definitive by any means; rather, our goal is to provide a solid beginning for continued, nuanced investigation. We welcome future research that builds from our conceptual platform: for example, research that investigates the differences between paced and self-paced courses in cMOOCs and xMOOCs severally; research that discerns the effect of pedagogical choice on crowd size in cMOOCs and xMOOCs; and, research that investigates the effects of crowd size on learning outcomes in cMOOCs and xMOOCs.

Finally, we welcome future research that implements the growing body of large dataset empirical work from the xMOOC domain (Brooks, *et al.*, 2015), and the rigorous experiments in this domain (Chudzicki, *et al.*, 2015; Lamb, *et al.*, 2015; Mullaney and Reich, 2015; Williams, *et al.*, 2015a; Williams, *et al.*, 2014) to assist in forming a generalizable crowd science (Prpić and Shukla, 2016), through meta-analyses, natural experiments, and formal models (Agrawal, *et al.*, 2014). Given the findings and implications of our work here, and given the unprecedented on-demand scale of human participation, the unprecedented on-demand speed and aggregation of human effort and knowledge, and the unprecedented on-demand access to human knowledge that we routinely see with both MOOC and crowdsourcing phenomena, using the fine-grained data from cMOOCs and xMOOCs to help assist in generalizing a science of crowds is now a very realistic possibility.

## About the authors

**John Prpić** is a Ph.D. candidate in the Faculty of Business Administration, Technology and Social Sciences at Lulea University of Technology in Sweden. John's research focuses on the organizational use of IT-mediated crowds in private and public sector settings, for the purposes of innovation, policy, public health, and collective intelligence.
E-mail: john [dot] prpic [at] ltu [dot] se

**James Melton** is an Associate Professor in the Department of Business Information Systems at Central Michigan University, where he teaches courses on social media, business communication, and intercultural communication. His research focuses on social media, particularly the disclosure- and privacy-related practices and perceptions of users in different cultural contexts.
E-mail: james [dot] melton [at] cmich [dot] edu

**Araz Taeihagh**\* (DPhil, Oxon) is an Assistant Professor of Public Policy at Singapore Management University. Since 2007, Taeihagh's research interest has been on the interface of technology and society on how to shape policies to accommodate new technologies and facilitate positive socio-technical transitions and how to use technology to enhance the policy process.
E-mail: araz [dot] taeihagh [at] new [dot] oxon [dot] org
\*Corresponding author.

**Terry Anderson** is a Professor Emeritus at the Centre for Distance Education at Athabasca University. His research focuses on interaction and social media use in higher education.
E-mail: terrya [at] athabascau [dot] ca

---

**Editorial history**

Received 10 August 2015; revised 15 November 2015; accepted 22 November 2015.

---